\documentclass[pageno]{jpaper}

\usepackage[normalem]{ulem}
\usepackage{algorithm}
\usepackage{algorithmic}
\usepackage{makecell}
\usepackage{array}

\usepackage[sort,nocompress]{cite}

\usepackage{amsmath}
\usepackage{amssymb}

\usepackage[dvipsnames]{xcolor}
\usepackage{wrapfig}
\usepackage{tabularx}
\usepackage{authblk}

\newcommand{\red}[1]{\textcolor{red}{#1}}

\usepackage{xspace}
\newcommand{\projectname}{Ripples\xspace}

\begin{document}

\title{Heterogeneity-Aware Asynchronous Decentralized Training}

\author[1]{Qinyi Luo \textsuperscript{\textasteriskcentered}}
\author[2]{Jiaao He \thanks{These two authors contributed equally. Jiaao He did this work during his internship at USC.}}
\author[1]{Youwei Zhuo}
\author[1]{Xuehai Qian}
\affil[1]{University of Southern California, CA, USA}
\affil[2]{Tsinghua University, Beijing, China}

\date{}
\maketitle

\thispagestyle{empty}

\begin{abstract}
Distributed deep learning training usually adopts All-Reduce as the synchronization mechanism for data parallel algorithms due to its high performance in homogeneous environment. However, its performance is bounded by the slowest worker among all workers, and is significantly 
slower in heterogeneous situations. AD-PSGD, a newly proposed synchronization method which provides numerically fast convergence and heterogeneity tolerance, suffers from deadlock issues and high synchronization overhead. 
Is it possible to get the best of 
both worlds --- designing a distributed training method
that has both high performance as All-Reduce in 
homogeneous environment and good heterogeneity tolerance
as AD-PSGD?

In this paper, we propose {\em \projectname},
a high-performance heterogeneity-aware 
asynchronous decentralized training approach.
We achieve the above goal
with intensive synchronization optimization,
emphasizing the interplay between
algorithm and system implementation. 
To reduce 
synchronization cost, we propose a novel communication primitive
Partial All-Reduce that allows 
a large group of workers to synchronize quickly.
To reduce
synchronization conflict, we propose static 
group scheduling in homogeneous environment
and simple techniques (Group Buffer and Group 
Division) to avoid conflicts with slightly 
reduced randomness. 
Our experiments show that in homogeneous environment, \projectname is $1.1\times$ faster than the state-of-the-art implementation of All-Reduce, $5.1\times$ faster than Parameter Server and $4.3\times$ faster than AD-PSGD. In a heterogeneous setting,
\projectname shows $2\times$ speedup over All-Reduce, and still obtains $3\times$ speedup over the 
Parameter Server baseline.


\end{abstract}

\section{Introduction}
\label{sec:introduction}
Deep learning is popular now. 
It has achieved phenomenal advancement in various fields including image recognition \cite{CVPR15:image}, speech processing \cite{Hinton:speech-recognition}, machine translation \cite{2016_translation_technologies}, gaming \cite{2016_game_go}, health care \cite{ml_healthcare_nature2018} and so on.
The key success of deep learning is the increasing size of models that can achieve
high accuracy. At the same time, it is difficult
to train the large and complex models. 
It is common that training a model may take hours
or even days~\cite{hpca2018_longtraining}. 
Therefore, it is crucial to accelerate training in the distributed manner to better prompt wider applications of deep learning.

In distributed training, multiple workers running
on
a number of 
compute nodes cooperatively train
a model with the help of communication between workers. 
The current widely used approach of distributed training is data parallelism~\cite{bennun2018demystifying_parallelisms}, in which each worker keeps a replica of the whole model, processes training samples independently, and synchronizes the parameters every iteration. 
{\em Parameter Server (PS)}~\cite{Li2014ParamServer} is the first approach to support distributed training by introducing a central node which manages 
one or more shared versions of the parameters of the whole model at PS.
More recently, {\em All-Reduce}~\cite{sergeev2018horovod}, an alternative distributed solution utilizing the advanced Ring All-Reduce algorithm~\cite{ring_allreduce1}, is shown to provide 
superior performance than PS~\cite{jia2018highly,sun2019optimizing,yamazaki2019yet,kurth2018exascale}. 
To fundamentally improve the scalability, 
the general
{\em decentralized training}~\cite{ArXiv_ADPSGD,NIPS2017_dPSGD,HOP,socc18_decentralized,HotCloud19_decentralized,NIPS18_pipeSGD_decentralized,pmlr-v80-tang18a_decentralized,NIPS2018CommunicationCF_decentralized} also received
intensive research interests. 
It has been recently theoretically shown for the first time that decentralized algorithms can outperform centralized ones \cite{NIPS2017_dPSGD}.
While PS and All-Reduce are both special
cases of the decentralized method, 
a general decentralized training scheme 
can use an {\em arbitrary} communication graph
with spectral gap, doubly stochastic averaging and
independence properties\cite{ArXiv_ADPSGD} to 
specify point-to-point communication between 
workers.



The first key problem of distributed learning is 
the {\em intensive communication}
among workers. 
During execution, gradients or parameter updates 
are transferred between workers in different 
nodes to achieve the eventually trained model. 
In PS, all workers need to communicate
with the parameter servers ---
easily causing communication bottleneck
even if the number of workers is relatively small. 
In All-Reduce, the communication is 
more evenly distributed among all workers, 
since it logically implements the 
all-to-all communication, the amount of 
parameters transferred is still high. 
More importantly, to hide communication latency, 
All-Reduce uses delicate pipelined operations among all workers.
It makes this solution vulnerable to 
system heterogeneity, a concept that means the performance of 
different nodes (workers) and 
the speed of different communication links
are different. 
Specifically, because All-Reduce requires global synchronization in every step, its performance is strongly bounded by the slowest worker, thereby
cannot tolerate heterogeneity well. 
We believe that {\em heterogeneity} is the second 
key challenge of distributed training.



To tolerate heterogeneity, both system and 
algorithm techniques have been proposed. 
At system level, backup worker~\cite{ICLR2016_backup_workers} and 
bounded staleness~\cite{NIPS2013_SSP} have been shown 
to be effective in mitigating the effects
of random worker slowdown in both PS~\cite{ICLR2016_backup_workers,DBLP:Tensorflow,NIPS2013_SSP,Petuum:SSP,SIGMOD16:SSP}
and decentralized training~\cite{HOP}.
However, if some workers experience severe 
and continuous slowdown, the benefits
of system solution are limited since the whole
system will eventually be dragged down 
by the slow workers or communication links. 
It motivates the more fundamental algorithm
level solutions. In particular, 
AD-PSGD~\cite{ArXiv_ADPSGD} probabilistically reduces
the effects of heterogeneity with randomized
communication.
In an additional synchronization thread, each worker randomly
selects one worker to average parameters between 
the two and {\em atomically} update both versions.
Moreover, the workers need to wait for the current synchronization
to finish before starting another, no matter if it actively initiates
a synchronization or is passively selected by another worker.
While the slow workers inevitably have
staler parameters and will drag down
others' progress, it will only happen if they
happen to be selected. 
Unfortunately, the implementation in ~\cite{ArXiv_ADPSGD} only supports a certain type of communication graphs and suffers from deadlock otherwise. 
More importantly, the parameter update protocol
in AD-PSGD incurs significant synchronization
overhead to ensure atomicity.

\begin{figure}
    \centering
    \includegraphics[width=0.48\textwidth]{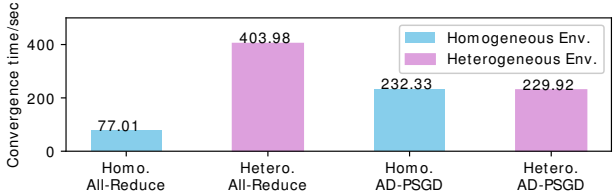}
    \caption{A comparison between All-Reduce~\cite{sergeev2018horovod} and AD-PSGD~\cite{ArXiv_ADPSGD} in homogeneous (Homo) environment and heterogeneous (Hetero) environment.}
    \label{fig:motivating_res}
\end{figure}

Figure~\ref{fig:motivating_res} shows the training 
performance\footnote{Defined as the time for training loss to reach $0.32$} of VGG-16 model over CIFAR-10 dataset, of All-Reduce~\cite{sergeev2018horovod} and 
AD-PSGD on $4$ GTX nodes running $16$ GPUs as $16$ workers in total in homogeneous and heterogeneous\footnote{The heterogeneous setting is that one worker is slowed down by $5$ times.} execution environment.
In Figure~\ref{fig:motivating_res}, we see 
AD-PSGD's excellent ability to tolerate heterogeneity --- $1.75$ times faster than 
All-Reduce.
However, the figure also
shows that All-Reduce is much faster ($3.02\times$) than AD-PSGD
in homogeneous environment.
Thus, the {\bf open question} is whether 
it is possible to improve 
AD-PSGD so that its performance is {\em comparable
to All-Reduce in a homogeneous environment
while still maintaining superior ability
to tolerate heterogeneity}?



In this paper, we propose {\em \projectname},
a high-performance heterogeneity-aware 
asynchronous decentralized training approach.
Compared to the state-of-the-art solutions, 
\projectname gets the best of both worlds:
it achieves better performance than All-Reduce
in homogeneous environment and significantly 
outperforms AD-PSGD in both 
homogeneous and heterogeneous environments. 
We achieve this almost ideal solution 
with intensive synchronization optimization,
emphasizing the interplay between
algorithm and system implementation. 
To {\em reduce 
synchronization cost}, we propose a novel communication primitive,
{\em Partial All-Reduce}, that allows 
a large group of workers to synchronize quickly.
To {\em reduce
synchronization conflict}, we propose static 
group scheduling in homogeneous environment
and simple but smart techniques (Group Buffer and Group 
Division) to avoid conflicts with slightly 
reduced randomness.




We perform experiments on Maverick2 cluster of TACC Super Computer. We train a common model VGG-16 on CIFAR-10 dataset to look deeply into different algorithms. We also train a large model, ResNet-50, on a large dataset, ImageNet, to validate the optimizations. Our experiments show that in homogeneous environment, \projectname is $1.1\times$ faster than the state-of-the-art implementation of All-Reduce, $5.1\times$ faster than Parameter Server and $4.3\times$ faster than AD-PSGD. In a heterogeneous setting,
\projectname shows $4.4\times$ speedup over All-Reduce, and also obtains $3.5\times$ speedup over the 
Parameter Server baseline.


\section{Background and Motivation}
\label{sec:background}

\subsection{Distributed Training}

In distributed training, a single model is trained collaboratively by multiple workers, which 
run in distributed compute nodes. 
Training is most commonly accomplished with Stochastic Gradient Descent (SGD), which is an iterative algorithm that reaches the minimum of the loss function by continuously applying approximate gradients computed over randomly selected data samples. In each iteration, there are typically three steps: (1) randomly select samples from the data set; (2) compute gradients based on the selected data; and (3) apply gradients to the model parameters.

There are a number of schemes to achieve 
parallelism among multiple workers in 
distributed training: 
data parallelism~\cite{sergeev2018horovod,sun2019optimizing}, model parallelism~\cite{cots_hpc}, hybrid parallelism~\cite{zhihao_hybrid,nyu_tofu}, and pipeline parallelism~\cite{msr_pipline}. 
Among them, data parallelism 
can be easily deployed
without significant efficiency loss compared with other models. 
Thus, it is supported by many popular machine
learning frameworks such as TensorFlow\cite{tensorflow2015-whitepaper}, MXNet\cite{DBLP:journals/corr/ChenLLLWWXXZZ15} and PyTorch\cite{Patarasuk:2009:BOA:1482176.1482266}.
Recent papers~\cite{zhihao_hybrid,nyu_tofu} discussed
the trade-offs between data parallelism and model
parallelism and proposed the hybrid approach. 
Due to space limit, we do not discuss other 
approaches in detail. 
Due to the 
popularity of data parallelism and the 
unresolved open problems, we focus on
this model in this paper.

In {\em data parallelism}, each worker consumes training data independently and computes gradients based on its own selected data. The gradients obtained by
distributed workers are then gathered and applied to model parameters during {\em synchronization}, 
and the updated model is subsequently used in the next iteration. Synchronization is both an essential part of parallelizing SGD and a critical factor in determining the training performance. 

\subsection{Existing Synchronization Approaches}

There are three main categories of approaches to performing synchronization in data parallelism: Parameter Servers (PS), All-Reduce, and generalized decentralized approaches.

Training with PS involves using one or more central nodes called \textit{Parameter Servers} that gather gradients from all workers and also send back the updated model to the workers. This straightforward approach enables relatively easy management 
of the training process.
However, PS has limited scalability 
due to the communication bottlenecks 
at Parameter Servers. 
Parameter Hub\cite{DBLP:journals/corr/abs-1805-07891} provides a new approach to remove the bottleneck of communication by introducing a new
network device to work as Parameter Server.
While promising, it requires special hardware 
supports that do not exist in common 
distributed environment (e.g., Amazon AWS). 


In contrast to PS, All-Reduce replaces the use of central nodes with carefully scheduled global communication to achieve better parallelism. 
The state-of-the-art solutions~\cite{sergeev2018horovod,sun2019optimizing,kurth2018exascale} leverage
Ring All-Reduce \cite{Patarasuk:2009:BOA:1482176.1482266}, the 
advanced all-reduce algorithm that effectively
utilizes the bandwidth between computation devices.
Specifically, workers are organized 
as a ring, and gradients are divided into chunks and passed over the ring in a parallel manner.
Different chunks of gradients are first
accumulated to different workers, which are
then broadcast to all workers in a parallel manner.
This algorithm achieves ideal parallelism within the theoretical upper bound. 
Another algorithm, Hierarchical All-Reduce~\cite{choblueconnectHierarchy, kurth2018exascale}, has been successfully scaled up to $4560$ nodes with $27360$ GPUs.
Utilizing All-Reduce algorithms based on MPIs\cite{mpistd,gabriel04:_open_mpi,intelmpi} and NCCL\cite{DBLP:journals/corr/ChenLLLWWXXZZ15}, Horovod~\cite{sergeev2018horovod} enables high-performance data parallelism and is proved to be effective and efficient --- based on All-Reduce algorithms and high performance implementations, researchers were able to use the fastest supercomputer , Summit\cite{summit}, to train a deep learning model in exascale\cite{kurth2018exascale}. 



Recently, the general decentralized approaches 
allow the point-to-point communication between workers
by specifying a communication graph. 
Both PS and All-Reduce can be considered as 
special case of the communication graph. 
Two main algorithms proposed so far are Decentralized Parallel SGD (D-PSGD)~\cite{NIPS2017_dPSGD} and Asynchronous D-PSGD (AD-PSGD)~\cite{ArXiv_ADPSGD}. In D-PSGD, 
every worker has its own version of parameters, and  only synchronizes with its neighbors in the graph. As training proceeds, local information at a worker propagates along edges of the communication graph and gradually reaches every other worker, and thus models at different workers converge collaboratively to the same optimal point. The convergence rate has been proved to be similar to that of PS and All-Reduce~\cite{NIPS2017_dPSGD}. Like All-Reduce, D-PSGD does not suffer from communication bottleneck. However, it relies on a fixed communication topology, which may be susceptible to heterogeneity (more discussion in Section~\ref{sec:challenges}).


To tolerate heterogeneity, AD-PSGD~\cite{ArXiv_ADPSGD} introduces a random communication mechanism on top of D-PSGD.
Instead of synchronizing with all the neighbors specified by the communication graph, a worker randomly selects a single neighbor, and performs an {\em atomic model averaging} with the neighbor, regardless of whether they are in the same iteration or not. 
While the slow workers inevitably have
staler parameters and will affect
the training of the global model, 
it will not block the progress of other workers unless it is selected, which happens only occasionally.



\subsection{Challenges and Problems}
\label{sec:challenges}


\noindent{\bf Communication} With the continuously increasing compute
capability (e.g., GPUs), communication has 
become more important and the focus of 
recent optimizations. 
The communication bottleneck
in PS has been eliminated by approaches based on Ring 
All-Reduce, but the latter's strongly synchronized communication 
pattern has lower heterogeneity tolerance. 
The generalized decentralized training
captures both schemes and enables more 
optimization opportunities. 



\noindent{\bf Heterogeneity}
With the communication problem largely mitigated,
performance degradation in the heterogeneous 
distributed environment becomes a major challenge. 
It is also known as the 
straggler problem, and occurs due to 
the performance difference among workers and 
the discrepancy or fluctuations of communication
speed and bandwidth. 
Heterogeneity is pervasive and can be caused
by multiple reasons such as resource sharing 
in data center, paging, caching and hardware faults.
The trend of heterogeneity and the ``long tail effects'' have been also discussed and confirmed
in other recent works~\cite{ICLR2016_backup_workers, Dean2013_hetero, SIGMOD2017_het,NSDI2017_Gaia,ArXiv_ADPSGD}.
A number of countermeasures for different 
synchronization schemes have been proposed, such as
asynchronous execution~\cite{NIPS2011_hogwild}, 
bounded staleness~\cite{NIPS2013_SSP}, 
backup workers~\cite{ICLR2016_backup_workers}, 
adjusting the learning rate of stale gradients~\cite{SIGMOD2017_het}, sending accumulated gradients over bandwidth-scarce links when they reach a significance threshold~\cite{NSDI2017_Gaia}, etc. 
Unfortunately, these techniques are mostly applicable for PS and 
decentralized training. 

For All-Reduce, with the delicate communication schedule, it is difficult to apply these ideas 
--- making it inherently vulnerable to heterogeneity.
From the computation aspect, a global barrier is introduced by the All-Reduce operation, so the throughput of computation is determined by the slowest worker in the cluster. 
From the communication aspect, although Ring All-Reduce algorithm is ideal in theory, the speed of sending chunks
along the ring is bounded by the edge with the slowest connection.

Considering the delicacy of All-Reduce, and due to the well-known limits of PS, tolerating 
heterogeneity in decentralized approach 
is particularly important. 
Recent work Hop~\cite{HOP} presented the first detailed
distributed protocol to support general
decentralized training~\cite{NIPS2017_dPSGD} with backup worker 
and bounded staleness to tolerate random slowdown. 
Although the results are promising, the 
proposed methods are essentially {\em system} techniques
to mitigate the effects of heterogeneity. 
The alternative way is {\em algorithmic} technique, with
AD-PSGD~\cite{ArXiv_ADPSGD} as an excellent example.
While AD-PSGD is both communication-efficient and
tolerates heterogeneity well, the atomic
model averaging step poses a key challenge of synchronization. 

\noindent{\bf Synchronization Conflict}
The atomic model averaging requires that two 
model averaging operations 
are {\em serialized}
if they involve the same worker. 
This requirement is to ensure fast 
convergence,
and more relaxed semantic will increase the 
mutual influence of model 
updates from different workers
--- making the global trained model more
vulnerable to ``staler'' updates.
Note that the problem is different from 
the synchronization relaxation in HOGWILD!~\cite{NIPS2011_hogwild}, where conflict happens when two workers try to update the same shared parameter. Conflict is expected to be rare, since HOGWILD! requires the cost function to be ``sparse'' and separable. In the algorithm, workers only update a small fraction of the parameters in each iteration, and the sparsity ensures that updates from different workers rarely involve the same parameter. Therefore, the algorithm can still converge even without any locks. However, in AD-PSGD, the conflict is of a different nature and is expected to be frequent, because every worker can initiate model averaging and it is likely that 2 of them end up choosing the same worker. 

\begin{figure}[h]
    \centering
    \begin{tabular}{m{.23\textwidth}m{.23\textwidth}}
    \includegraphics[width=.23\textwidth]{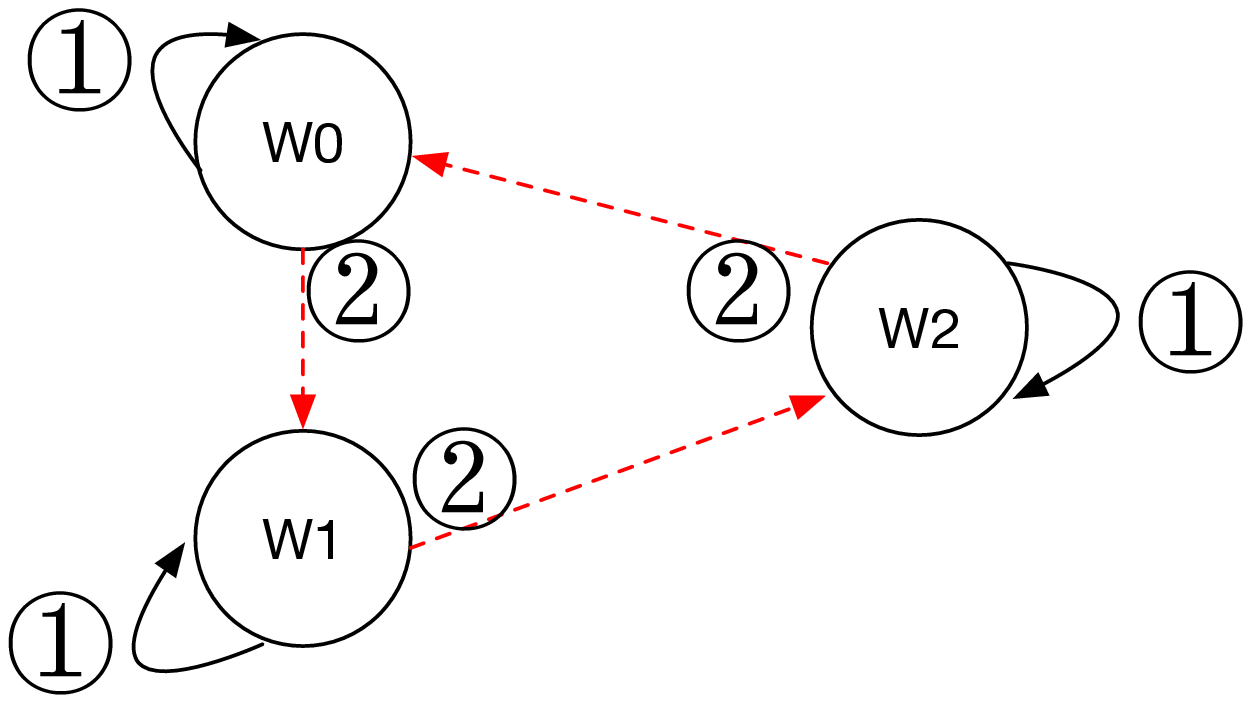}
    \footnotesize
    (a) An example deadlock happens when all workers first lock themselves (\textcircled{1}), and then try to lock their neighbors in a cycle (\textcircled{2}), which blocks forever.
    &
    \includegraphics[width=.23\textwidth]{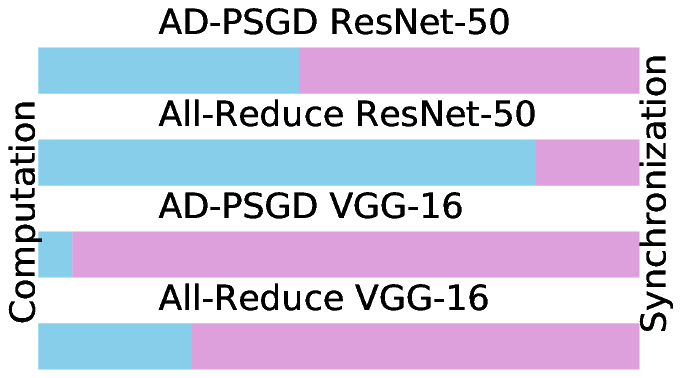}
    \footnotesize
    (b) Computation and synchronization ratio of different algorithms on different tasks
    \end{tabular}
    \vspace{-2mm}
    \caption{Synchronization issues of AD-PSGD}
    \label{fig:ratio}
\end{figure}

To ensure atomic model averaging and avoid deadlock as exemplified in Figure \ref{fig:ratio}(a), AD-PSGD divides the workers into 2 sets --- active set and passive set, and requires that edges in the communication graph only exist between the two sets, i.e., neighbors of active workers can only be passive workers, and vice versa. This division is only possible when the communication graph is bipartite. In the implementation, only active workers are allowed to initiate model averaging, while passive workers 
can only respond.
This is slightly different from the algorithm, in which every worker can initiate averaging.
When an active worker needs to synchronize, it sends its model to the selected neighbor and blocks until it gets a response.
Possible violation of atomicity can only happen when 2 active workers select the same passive worker, and it can be avoided by letting the passive worker deal with the requests one by one. Note that this scheme will incur deadlock if all workers are allowed to initiate model averaging or if the graph is not bipartite. 

Besides the restriction of the communication graph
between workers, the synchronization overhead
is a more crucial problem in a distributed 
environment. 
When training VGG-16 model over CIFAR-10, and ResNet-50 model over ImageNet using AD-PSGD on $16$ GPUs, Figure \ref{fig:ratio}(b) shows that more than $90\%$ of the time can be spent on synchronization in AD-PSGD.
This is measured by comparing per iteration time of  workers without synchronization (i.e., skip the synchronization operation to see the actual time of computation) and workers with the synchronization enabled.

\section{Partial All-Reduce}
\label{sec:preduce}
Based on the results in Section \ref{sec:challenges}, 
we mainly focus on the 
synchronization challenge for decentralized
training. 
This section first presents a deep analysis
of AD-PSGD which motivates our key contribution of 
Partial All-Reduce primitive. 

\subsection{AD-PSGD Insights}

\begin{figure}
\centering
\small
 \begin{algorithmic}[1]
 \REQUIRE A set of workers represented as nodes $V$ in a graph and the connection among them are represented by an adjacency matrix $W$
 \FOR{worker $i \in V$}
 \STATE Initialize model weights $x_i$
 \WHILE{not reached convergence}
 \STATE Step 1. Read the local model $x_{i}$ from memory
 \STATE Step 2. Compute gradients over randomly selected samples $\xi_{i}$, and update weights: $x_{i} \leftarrow x'_{i}-\eta_k \cdot \nabla F(x_{i};\xi_{i})$
 \STATE Step 3. Randomly select a neighbor $j$
 \STATE Step 4. {\bf Atomically} average weights with the selected neighbor and update the local model as well as the selected neighbor's model: $x_{i},x_{j} \leftarrow \frac{1}{2} (x_{i} + x_{j})$
 \ENDWHILE
 \ENDFOR
 \end{algorithmic}
\footnotesize
\emph{Notes:} $x'_{i}$ may be different from $x_{i}$ since it may have been modified by other workers in their averaging step (i.e., step 4). To ensure the correctness of execution, it is crucial to implement the averaging step atomically with certain locking mechanisms.
\caption{AD-PSGD Algorithm }
\label{alg0}
\end{figure}

AD-PSGD algorithm is shown in Figure \ref{alg0}. 
Similar to traditional training such as PS and All-Reduce, 
in one iteration, 
it computes gradients first, and then performs synchronization; the difference is that it only synchronizes with a random selected neighbor, instead of all the workers. 
Therefore, the global barrier is removed, 
enabling higher training throughput 
and better heterogeneity tolerance.

In AD-PSGD, each worker $i$ has a local version of 
parameters,
which can be seen as a single concatenated vector $x_i$, as the shapes do not matter in synchronization. Concatenating all the weight vectors together, they can be represented as a matrix $X = [x_1 x_2 \dots x_n] \in \mathbb{R}^{N\times n}$ where $N$ is the total size of weights in the model, and $n$ is the number of workers.

In this formalization, one iteration
in a worker in AD-PSGD algorithm 
can be seen as one update to $X$. Formally, it can be represented as:
$ X_{k+1} = X_k W_k - \gamma \partial_g(\hat{X_k};\xi_k^{i},i) $.
Here, $\partial_g(\hat{X_k};\xi_k^{i},i)$ is the update to $x_i$ according to gradient computation based on a random worker $i$, the previous version of $\hat{x_i}$, and a random subset of the training samples $\xi_k^{i}$. $W_k$ is a {\em synchronization matrix} that represents the process of model averaging: $x_i, x_j \leftarrow \frac{1}{2} (x_i + x_j)$. 

\begin{figure}
    \centering
    \vspace{-1mm}
    \includegraphics[width=0.4\textwidth]{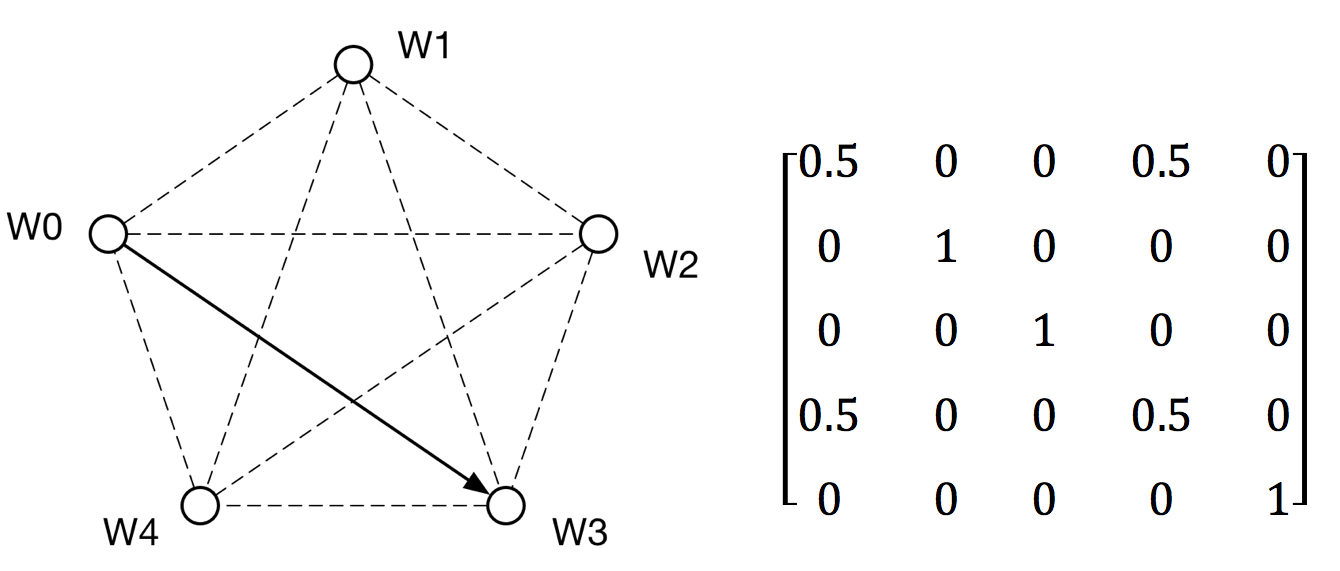}
    \vspace{-1mm}
    \caption{Synchronization in AD-PSGD }
    \vspace{-2mm}
    \label{fig:adpsgd_1pair}
\end{figure}

Figure \ref{fig:adpsgd_1pair} shows an example of $W_k$, in which worker $0$ performs a synchronization with worker $3$. More generally, for an update between worker $i$ and worker $j$, the non-zero entries of matrix $W_k$ are:
$W^k_{i,i} = W^k_{i,j} = W^k_{j,i} = W^k_{j,j} = 0.5$,
$W^k_{u,u} = 1, \forall u \neq i, j$.

In AD-PSGD, a {\em conflict} happens when 
two workers $i, j$ both select another worker $u$ for synchronization. In order to keep the atomic property of weight updating, 
the two operations need to be serialized. 
In matrix formalization, 
assume that $W_k$ represents the synchronization between $i$ and $u$, $W_{k+1}$ represents the synchronization between $j$ and $u$. Ignoring the gradient entry in the update, the updated weight $X_{k+2}$ can be represented 
as:
$ X_{k+2} = X_{k+1} W_{k+1} =  (X_k W_k) W_{k+1} = X_k (W_k W_{k+1})$.

\begin{figure}
    \centering
    \includegraphics[width=0.4\textwidth]{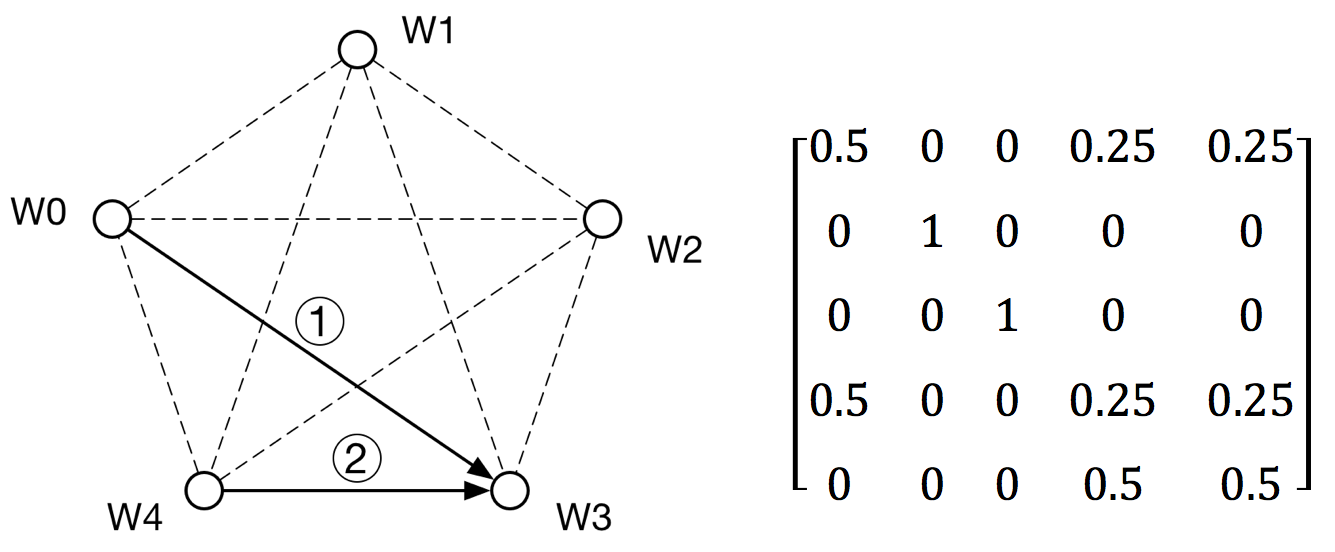}
    \caption{Conflict between two pairs of workers}
    \vspace{-2mm}
    \label{fig:adpsgd_2pairs}
\end{figure}

Figure \ref{fig:adpsgd_2pairs} shows an example of two workers $w_0$ and $w_4$ requiring synchronization 
with the same worker $w_3$ ($i=0, j=4, u=3$).
The matrix on the right shows the production of $W_k$ and $W_{k+1}$ as a {\em fused synchronization matrix} $W_{fused} = W_k W_{k+1}$, which shows the final update over all the weights.

We can observe that the production is 
{\em commutative} in AD-PSGD --- $W_k$ and $W_{k+1}$ can be exchanged (not mathematically but logically). It is because the order
of synchronization is determined by the order of 
getting a lock, which is a completely random.
Based on the atomicity requirement, 
the key insight is that in AD-PSGD, although the two synchronizations can be mathematically fused, they have to be executed {\em sequentially}. 

\subsection{Partial All-Reduce and Group Fusion}

We propose {\em Group Fusion} --- fusing multiple synchronizations approximately into one 
with reduced synchronization cost. 
In the precise fused synchronization, 
according to the $W_{fused}$ matrix,
several workers update their weights to a certain linear
combination of the weights of each worker in the group. 


Next, we seek proper approximation of the 
fused synchronization to achieve 
efficient implementation. 
Our goal is to leverage 
Ring All-Reduce, the high-performance algorithm that can
compute the mean of several copies of weights in $O(N)$ time. 
We cannot directly use All-Reduce to execute the 
synchronization among the three workers in 
Figure \ref{fig:adpsgd_2pairs}.
It is because All-Reduce produces the same update for each worker, which is different from the outcome produced by multiplying a sequence of synchronization
matrices in a certain order (on the right of 
Figure \ref{fig:adpsgd_2pairs}).

Thanks to the commutative property of $W_k$'s, 
our {\bf key idea} is to {\em slightly relax the entries in $W_{fused}$
to leverage All-Reduce to perform the synchronization specified by $W_{fused}$}.
Generally, assume that there is a group of workers $G=\{w_1,w_2,\dots,w_k\}$ that perform a single fused synchronization together, $W_{fused}$ involves modifying the weights of all the workers in $G$. The 
{\em $W_{fused}$ with approximation} is defined as $F^G$, which contains the following non-zero entries:
$ F^G_{i,j} = \frac{1}{|G|}, \forall i, j \in G $,
$ F^G_{u,u} = 1, \forall u \notin G $.

\begin{figure}
    \centering
    \includegraphics[width=0.4\textwidth]{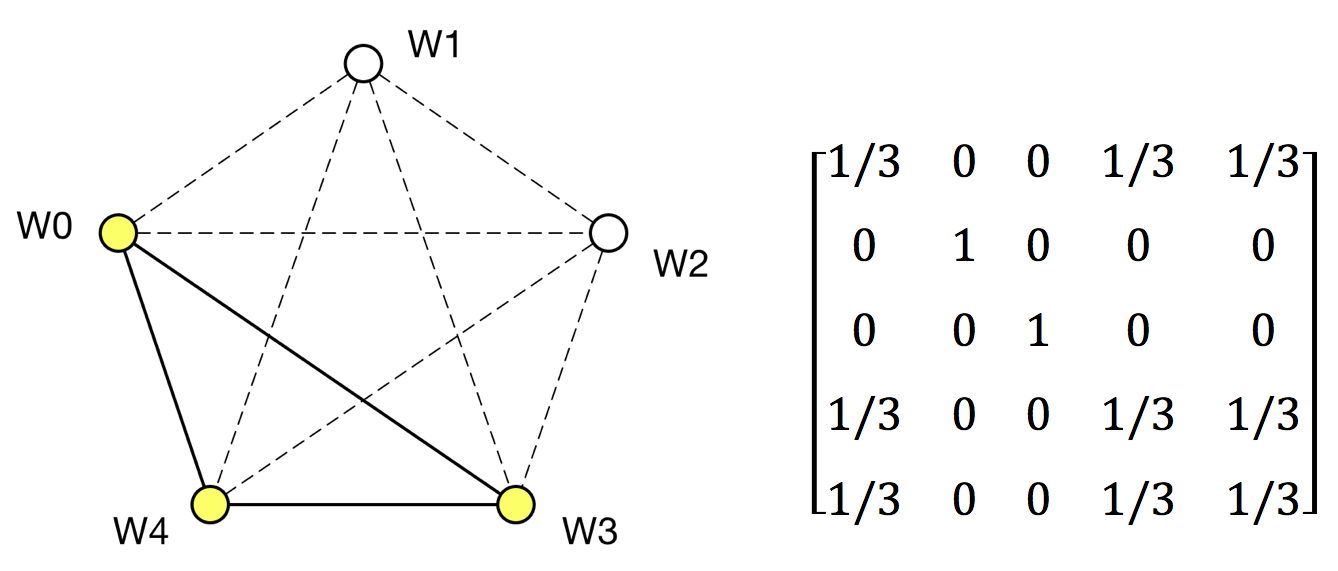}
    \caption{Synchronization with Partial All-Reduce}
    \vspace{-2mm}
    \label{fig:adpsgd_preduce}
\end{figure}

Figure \ref{fig:adpsgd_preduce} shows an example of $F^G$ among worker $0, 3, 4$ with the modified $W_{fused}$. Although the example only involves $3$ workers, 
the group can contain an arbitrary number of workers.

\begin{figure}
\centering
\small
 \begin{algorithmic}[1]
 \REQUIRE A set of worker represented as nodes $V$ in a graph and their connection represented by a weighted adjacency matrix $W$
 \FOR{worker $i \in V$}
 \STATE Initialize model parameters $x_i$
 \WHILE{not reached convergence}
 \STATE Step 1. Read the local model $x_{i}$ from memory
 \STATE Step 2. Compute gradients over randomly selected samples $\xi_{i}$, and update parameters: $x_{i} \leftarrow x_{i}-\eta_k \cdot \nabla F(x_{i};\xi_{i})$
 \STATE Step 3. Randomly generate a group $G$ {\em including $i$}.
 \STATE Step 4. Atomically average parameters in group $G$ using P-Reduce:
 \STATE $\bar{x}_G= \frac{1}{|G|} \sum_{\forall g \in G} x_g$
 \STATE $x_g \leftarrow \bar{x}_G, \forall g \in G$
 \ENDWHILE
 \ENDFOR
 \end{algorithmic}
\footnotesize
\caption{Proposed algorithm using P-Reduce}
\vspace{-3mm}
\label{alg1}
\end{figure}

Applying $F^G$ is equivalent to {\em performing All-Reduce in the group $G$}. We define this operation as {\em Partial All-Reduce or P-Reduce} to distinguish our algorithm from the conventional All-Reduce in deep learning training that performs All-Reduce among {\em all} workers. 
Based on P-Reduce, we present a formal description of the new algorithm in Figure \ref{alg1}.

Compared to the original AD-PSGD algorithm, there are 
two key differences. 
First, in Step 3, each worker can randomly generate a group
that may be larger than 2, as long as it contains itself, $w_i$.
The group in AD-PSGD of size 2 (one worker randomly selects
a neighbor) becomes a special case.
It essentially enlarges the {\em unit of synchronization}
to groups of any size. 
Larger groups have two implications:
(1) potentially enable fast propagation 
of model parameter updates among workers, speeding up 
convergence; and 
(2) increase the chance of conflicts. 
Thus the new algorithm allows the system to explore 
such a trade-off. 
The second difference from AD-PSGD is that the synchronization operation is performed by 
the new primitive P-Reduce involving the workers in the group, instead of using individual messages among workers. 
This directly reduces the cost of synchronization. 

Although group fusion 
inspired us to propose the idea of P-Reduce, 
the algorithm in 
Figure \ref{alg1} does {\em not} need to fuse
groups during execution. In fact, the effects of 
fusing two groups of size 2 in AD-PSGD is reflected 
as generating group of arbitrary size in Step 3
of Figure \ref{alg1}.
As a result, \projectname only needs to deal with 
group generation but not group fusion. 
The system still needs to satisfy the atomicity 
requirement. 
If two $G$'s do not share common workers, the two 
non-conflicting $F^G$'s can be executed concurrently. In an unrealistic but ideal situation, applying all the $F^G$'s 
should not introduce any conflict. 
Compared to All-Reduce, P-Reduce retains the efficient 
implementation while avoiding 
the global barrier. 

\subsection{Convergence Property Analysis}

To guarantee that models at different workers converge to the same point, three requirements for $W_k$ are proposed in AD-PSGD\cite{ArXiv_ADPSGD}. 
In the following, we show that although $F^G$
is not exactly the same as the result of
multiplying a sequence of synchronization
matrices in a certain order,
our definition of $F^G$ satisfies all three
convergence properties as AD-PSGD does.


{\bf Doubly stochastic averaging} $W_k$ is doubly stochastic for all $k$. The sum of each row and each column equals to $1$ in both $W_k$ and $F^G_k$.

{\bf Spectral gap} There exists a $\rho \in [0, 1) $, such that:
$max\{|\lambda_2 (\mathbb{E} [W_k^T W_k])|, |\lambda_n (\mathbb{E} [W_k^T W_k])|\} \leq \rho, \forall k$.
Basically, $(F^G)^T F^G = F^G$. And $\mathbb{E} [F^G]$ can be regarded as a Markov Transition Matrix. According to the Expander Graph Theory\cite{Expander}, the spectral gap condition is fulfilled if the corresponding graph of random walk is connected. That means the update on any worker can be passed through several groups to the whole graph. When creating the group generation methods in the following section, this property is always kept in our mind to guarantee the convergence property.

{\bf Dependence of random variables} $W_k$ is a random variable dependent on $i_k$\footnote{$i_k$ is the worker initiating the synchronization.}, but independent on $\xi_k$ and $k$. Up to now, the only requirement on the generated group $G_k$ is that it should contain the initial worker $i_k$. Theoretically, it is generated randomly without any connection to $k$ or $\xi_k$. Therefore, this condition is fulfilled.

\section{Group Generation and Conflict Detection}
\label{sec:gg}
With P-Reduce, 
a group of workers becomes
the basic unit of synchronization procedure. 
As a type of collective operation, 
all workers in the group need to call P-Reduce function.
It means that 
all group members should have the same 
group information to initiate the P-Reduce. 
It is non-trivial to obtain the consistent group among 
all workers inside the group. 
This section discusses how to generate the groups and
serialize conflicting groups. 

\subsection{Group Generator}

In Figure \ref{alg1}, each worker needs to randomly 
generate a group. This can be performed by 
each worker based on the communication graph
with randomly selected neighbors.
The workers in each group will collectively perform
P-Reduce. 
The system needs to ensure atomicity ---
P-Reduces of groups with overlapping workers selected
must be serialized. 
This can be implemented in either a centralized or 
distributed manner. 
In general, a distributed protocol involves
multiple rounds of communication and coordination
between workers. 
For simplicity, \projectname implements a 
centralized component. 
We can actually offload the group generation 
functionality from the workers to this component.
Thus, we call it {\em Group Generator (GG)}.
When a worker needs to perform a synchronization, 
it just needs to contact GG without any group information,
and then GG can select the group on behalf of the worker and
maintain the atomicity. 
In the following, we explain the protocol using an example.
We will find that the communications between workers and 
GG are only small messages, and 
do not introduce communication or scalability bottleneck.

\begin{figure}
    \centering
    \includegraphics[width=0.27\textwidth]{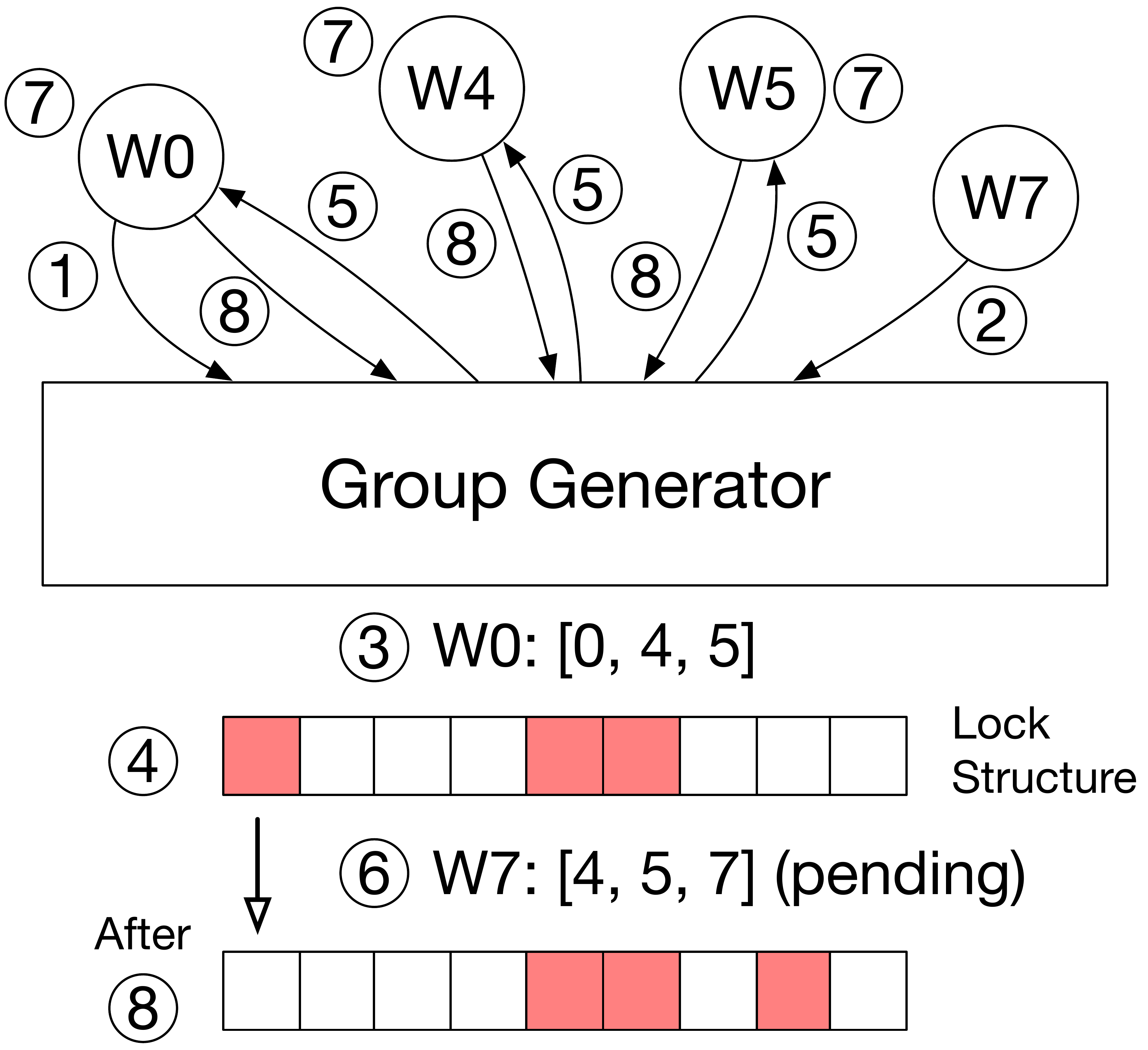}
    \caption{GG generates groups on behalf of workers}
    \vspace{-2mm}
    \label{fig:gg_lock}
\end{figure}

In Figure~\ref{fig:gg_lock}, we consider four 
workers $W_0$,$W_4$,$W_5$,$W_7$ among a total number of 
8 workers. 
In the beginning, $W_0$ and $W_7$ finish an iteration
and need to perform a synchronization. 
Instead of generating groups locally, they 
both send a synchronization request to GG, indicated
in \textcircled{1} and \textcircled{2}.
GG maintains the atomicity with a local lock vector
--- a bit vector indicating whether each worker is currently performing a P-Reduce. 
This vector is initialized as all 0s. 
Assume that there is no other synchronization 
being performed in the system, and GG receives the request
from $W_0$ first.
After that, GG randomly generates a group $[0,4,5]$ on behalf of 
$W_0$ (\textcircled{3}) and sets the corresponding
bits in the lock vector (\textcircled{4}). 
Then, GG notifies the workers $W_0$, $W_4$, and $W_5$ (\textcircled{5})
in the group so that they can collectively perform the 
P-Reduce. 
Later, GG receives the synchronization request from 
$W_7$ and randomly generates a group $[4,5,7]$.
Unfortunately, it is conflicting with the first group
due to the two overlapped workers $W_4$ and $W_5$, and 
needs to be serialized.
We can achieve this by simply blocking the group
$[4,5,7]$ and storing it in a pending group queue (\textcircled{6}).
In the meantime, $W_0$,$W_4$ and $W_5$ receive the 
notifications from GG and perform P-Reduce (\textcircled{7}). 
They also need to acknowledge GG to release the locks
(\textcircled{8}). 
After the locks for group $[0,4,5]$ are released in GG, 
the group $[4,5,7]$ can be performed after
setting the corresponding bits in the lock vector.

\subsection{Decentralized Static Scheduler}

As we have seen in the example in Figure~\ref{fig:gg_lock}, two overlapping groups need to be serialized to ensure atomicity, causing delay in the execution. We can eliminate the conflict by statically scheduling the groups to be non-overlapping.

\begin{figure}
    \centering
    \includegraphics[width=0.48\textwidth]{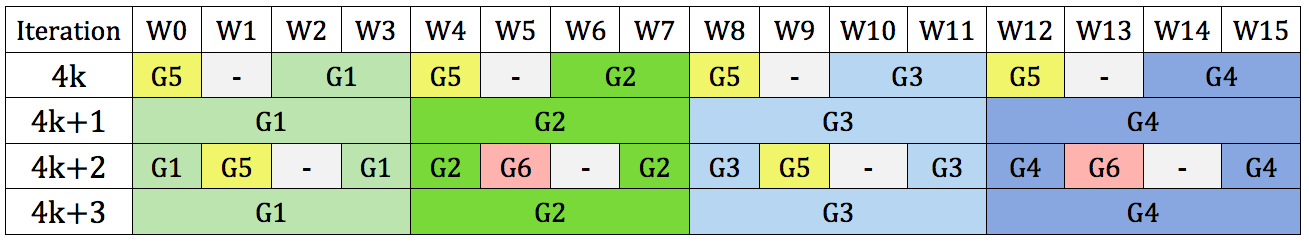}
    \caption{A conflict-free static scheduling strategy}
    \vspace{-2mm}
    \label{fig:schedule}
\end{figure}

We design a conflict-free schedule as shown in Figure~\ref{fig:schedule}. There are 16 workers in total, and the schedule is periodic with a cycle length of 4. Every row corresponds to an iteration, and colored blocks with group indices indicate the grouping of workers. For example, in the first row, $W_0$, $W_4$, $W_8$ and $W_{12}$ are all colored yellow with an index ``G1'', which means that these 4 workers are in the same group in the $(4k)$-th iteration, for any $k \in \mathbb{N}$. Group indices do not indicate the sequence of execution; in fact, groups in the same row are expected to execute concurrently. In addition, some workers do not participate in synchronization in certain iterations, and this is shown by gray blocks marked with a hyphen "-". For instance, $W_2$, $W_6$, $W_{10}$ and $W_{14}$ do not participate in any group in the $(4k+2)$-th iteration, for any $k \in \mathbb{N}$. Skipping synchronization can decrease the frequency of communication and thus shorten the training time. It is a technique that has been proved helpful in \cite{ppopp2019_periodic_update,Wang2018AdaptiveCS_periodic_update}.

To implement static scheduling, a naive way is to store the schedule table in the GG, and workers can access it by contacting the GG. Alternatively, we can store the table inside each worker, saving a round trip of communication between the worker and the GG. Since every worker has the same schedule table stored locally, a consistent view of the groups is naturally ensured.

\begin{figure}
    \centering
    \begin{tabularx}{.48\textwidth}{m{0.05\textwidth}|XXXX}
    \hline
    Phase & 
    {L.W. 0} & 
    {L.W. 1} & 
    {L.W. 2} & 
    {L.W. 3} \tabularnewline\hline
    
    0 & 
    \scriptsize 
    \cellcolor{yellow!25} Sync with L.W. 0s {\em on ALL NODES} &
    No sync & 
    \cellcolor{green!25} Sync with L.W. 3 & 
    \cellcolor{green!25} Sync with L.W. 2 \tabularnewline
    
    1 & \multicolumn{4}{c}{\cellcolor{blue!25} Sync L.W. 0-3} \tabularnewline
    
    2 &
    \cellcolor{green!25} Sync with L.W. 3 &
    \cellcolor{purple!25} \scriptsize Sync with L.W. 1 {\em on the opposite node on the ring} & 
    No sync & 
    \cellcolor{green!25} Sync with L.W. 0 \tabularnewline
    
    3 & \multicolumn{4}{c}{\cellcolor{blue!25} Sync L.W. 0-3} \tabularnewline\hline
    \end{tabularx}
    \footnotesize 
    {\emph Notes:} This table shows the rules that generate the schedule for $4$ workers running on {\em one node}. The rules are {\em the same for all 4 nodes}. L.W. $k$ stands for Local Worker $k$, the $k$-th worker on this node.
    The schedule has $4$ phases, each corresponds to one training step. It repeats itself after every 4 steps.
    \caption{An example of the static scheduling algorithm 
    }
    \vspace{-2mm}
    \label{fig:scheduler}
\end{figure}

In fact, storing a table is unnecessary, since the schedule is generated in a rule-based manner. For example, our previously proposed schedule is based on a worker's rank in its node. In an example where $4$ workers are on a node, the rule of scheduling is shown in Figure \ref{fig:scheduler}. 
In this way, a worker can simply call a local function $S$ to obtain its group in an iteration. The logic of $S$ guarantees that the schedule is consistent among all the workers, and a conflict-free static schedule is therefore enforced.

\subsection{Discussion: Random vs. Static}
Although static scheduling can ideally eliminate conflict and speed up execution, randomized group generation is more suitable for heterogeneous environment. 
We compare the different characteristics of
the two approaches below. 

Random GG is centralized, but it is different from Parameter Servers in that it
does not involve massive weight transfer. It only costs minor CPU and network resources compared with gradient accumulation or weight synchronization. In our experiment, it is found that GG can be put on a node together with workers 
without incurring any performance loss. 
However, in random GG, contacting the GG induces communication overhead, 
and conflicting groups need to be serialized, resulting in additional wait time.

On the contrary, GG implemented as a static scheduler has no communication latency. With a proper design of $S$, it can not only fully parallelize synchronization, but also utilize the architecture of the worker devices to accelerate every single P-Reduce operation. For example, it can schedule more intra-node synchronizations,
and reduce the number of large-scale inter-node synchronizations. However, the $S$ function is pseudo random, which breaks the strict convergence condition of AD-PSGD, although 
the resulting algorithm
still converges well in our experiments.  

When a certain worker is slower than others, the original AD-PSGD algorithm is able to tolerate the slowdown.
However, the static scheduler does not have such ability, as the schedule is in fact fixed. Synchronizations with the slow worker will slow down the whole training. 
As for random GG, the stragglers' effect can be largely ameliorated. Well-designed group generation strategy can ensure that at any time, most workers will be able to proceed without depending on the few slow workers,
thus relieving the slowdown problem. Also, slowdown detection and conflict avoidance mechanisms, which will be discussed in the following section, can be easily integrated into random GG,
making it
better adapt to heterogeneous environment.

\section{Smart Randomized Group Generation}
\label{sec:conflict}
The basic implementation of the scheduler 
in GG is to always randomly generate a group 
as specified in Step 3 of Figure~\ref{alg1}.
With the centralized GG,
our objective is to leverage the global and runtime
information to generate groups in a more 
intelligent manner to:
(1) avoid conflicts; and 
(2) embrace heterogeneity. 
 For example, a worker may have already been 
 assigned to several groups and thus have 
 several pending P-Reduces to perform. 
 If the worker is still selected to be included
 in a new group, then other workers will have to 
 wait for all the prior scheduled P-Reduces to finish.
 Similarly, when a slow worker is in a group, the whole group may be blocked by this worker.
Moreover, performing P-Reduce in different groups costs different time due to architecture factors. 
The group selection can even introduce 
architectural contentions on communication links. 
Based on the above insights,
we propose intelligent scheduling mechanisms for GG
to further improve performance. 


\subsection{Conflict Avoidance by Global Division}

An intuitive way of reducing conflict is to have a 
{\em Group Buffer (GB)} for each worker, which 
includes the ordered list of groups that include
the corresponding worker. 
When a group is formed, 
the group information is inserted in the GB
of all workers involved. 
The consensus group order can be easily ensured 
among all GBs since the GG, as a centralized structure,
generates groups in a serial manner. 
Based on GB, when GG receives a synchronization
request from a worker, it can first look up the 
worker's GB. If it is empty, a new group is 
generated for the worker; otherwise, the first existing group in the worker's GB will serve
as the selected group.

The main insight is that P-Reduce is 
a collective operation. So if $W_i$ initiates
a synchronization with $W_j$, i.e., $W_i$ and $W_j$
are in the same group, P-Reduce of this group
is only performed when $W_j$ also requests its
synchronization. Therefore, the simple mechanism
can avoid generating a new group for $W_j$ when 
it is already scheduled (and ready) to execute
a P-Reduce. 
However, with random group generation, nothing 
would prevent the selection of $W_j$
into a different group {\em not} initiated by $W_i$. 
In this case, the overlapping groups and the corresponding P-Reduce operations are still 
serialized.



\begin{figure}
    \centering
    \includegraphics[width=0.96\linewidth]{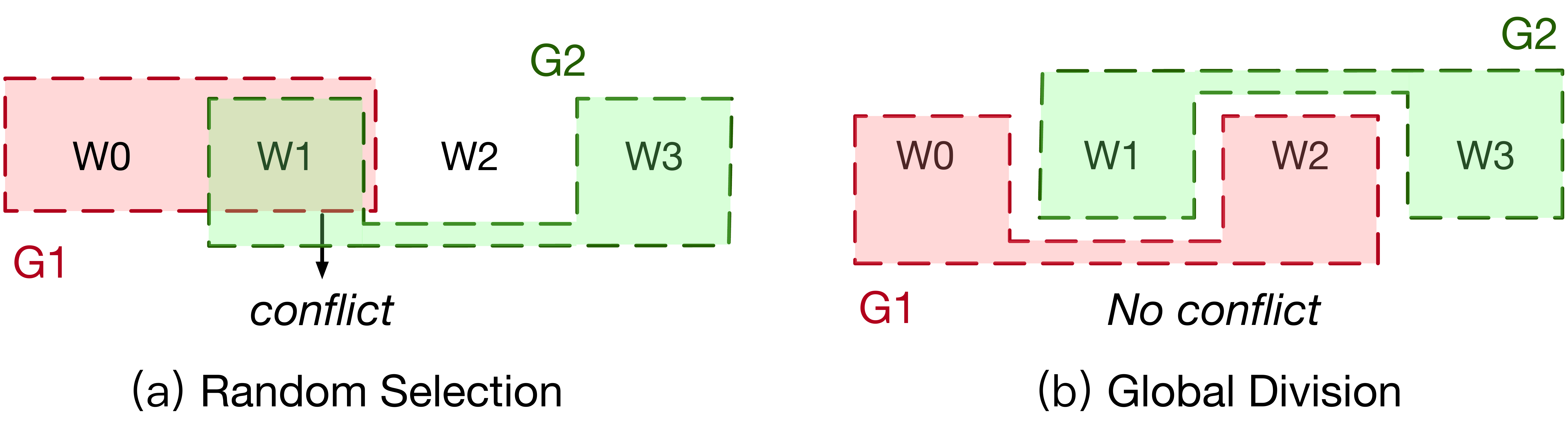} 
    \footnotesize
    {\emph Notes:} In random selection shown in (a), after $G_1$ is generated by request from $W_0$ and $W_1$ gets its group, no information is left to avoid the conflict that another request from $W_3$ may also generate a group including $W_1$. In GD shown in (b), two groups are both generated upon the first request. Therefore, the second request directly gets a conflict-free group from the buffer.
    \caption{An example of Global Division}
    \vspace{-2mm}
    \label{fig:global_devision}
\end{figure}

Inspired by the static scheduling,
we propose an operation called {\em Global Division (GD)} that
{\em divides all current workers with empty GBs
into several non-conflicting groups}.
A GD is called whenever a worker needs to generate
a group and its GB is empty. 
A simple example is shown in Figure~\ref{fig:global_devision}. In total we have 
4 workers and initially all GBs are empty. 
On the left, random selection shows a possible scenario without GD optimization.
The groups are randomly 
generated, so if $G_1$ initiated by 
$W_0$ includes $W_0$ and $W_1$,
another group $G_2$ initiated by $W_3$ can still 
include $W_1$ as the overlapping worker, 
thus introducing a conflict. 
On the right, with GD, when $W_0$ requests
a group, the GG will not only generate one for it,
i.e., $[W_0,W_2]$, but also randomly generate groups
for other workers, i.e., only $[W_1,W_3]$ 
in this example as there are only 4 workers. 
In this way, when later $W_3$ requests a group, 
GG will directly provide the non-conflicting
$[W_1,W_3]$ generated before. 

It is worth emphasizing two conditions. 
First, a GD only generates 
groups for the current ``idle''
workers (including the caller worker)
that are not assigned to any group. 
Thus, when a worker requests a group, 
it is possible to generate groups in the above 
manner for just a {\em subset} of workers. 
Second, a GD is only called 
when the initiator's GB is empty, otherwise
the first group in the initiator's GB will be 
returned. 

Indeed, the proposed schemes to avoid
conflict make the group generation not
fully random. 
However, we argue that the effects are not 
critical. For the first optimization 
based on GB, we only reuse the existing group
involving the worker who is requesting 
synchronization. 
This group is still generated in a fully 
random manner (if we do not use GD). 
For GD, essentially we generate a 
{\em random group partition} among all idle
workers together, which is triggered by 
the first worker in the set who initiates
a synchronization. 
So the difference is between randomly generating
each group and generating a random partition. 
We acknowledge that they are not the same but 
believe that our method does not significantly
damage the randomness. 
We leave the theoretical analysis as the future work.
However, based on the results shown in our 
evaluation, the ideas work very well in practice.


\subsection{Architecture-Aware Scheduling}

If the groups are randomly divided, multiple groups may all need to use the network bandwidth at the same time, causing congestion, which is not optimal in the perspective of architecture. In fact, All-Reduce is fast because it has a balanced utilization of different connections between different devices, such as Infiniband HCA cards, QPI paths\footnote{The Intel QuickPath Interconnect between CPU sockets within one node}, and PCIe slots.
To better utilize the bandwidth of different connections, we propose a new communication pattern called {\em Inter-Intra Synchronization} that can be naturally 
incorporated with GD. 
Here, a node, commonly running $4$ or $8$ workers,
are considered a unit. 
The scheme has an {\em Inter} and an {\em Intra} phase.


\begin{figure}
    \centering
    \includegraphics[width=0.96\linewidth]{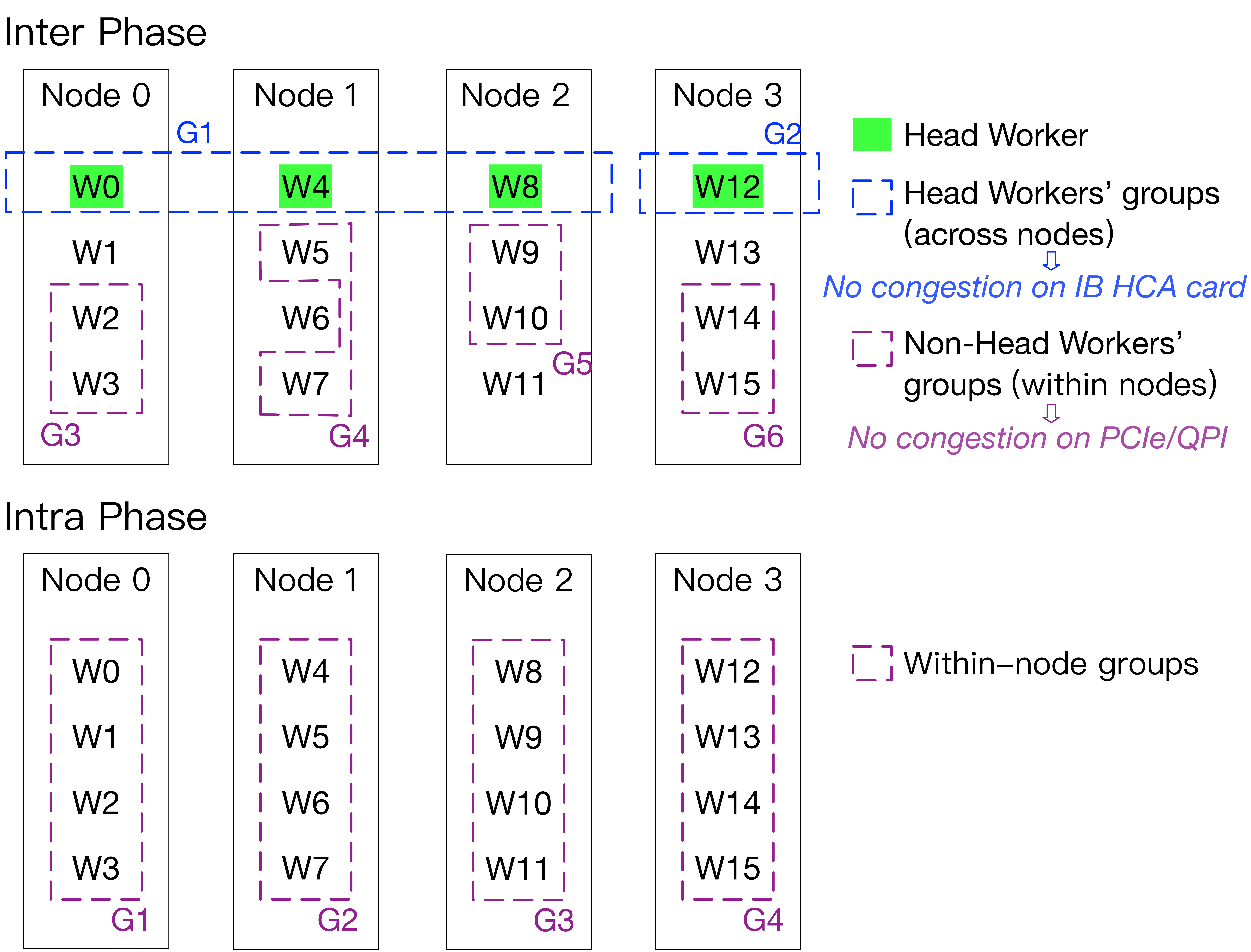} 
    \caption{An example of Inter-Intra Synchronization}
    \vspace{-1mm}
    \label{fig:inter_intra}
\end{figure}


\noindent{\bf \textit{Inter} phase} One worker on each node is selected as {\em Head Worker} of the node. All the {\em Head Workers} are randomly divided into several groups to synchronize in a inter-node manner. At the same time, the workers that are not {\em Head Worker} are randomly assigned to groups with only 
local workers in the same node. 
In this way, only the Head Worker can generate 
inter-node communication while the others only incur local communication, which can be carefully arranged  to avoid congestion on PCIe switches or QPI. 

\noindent {\bf \textit{Intra} phase} Workers within a node  synchronize with {\em all other local workers} collectively. In another word, it involves a P-Reduce among all the workers in the same node, without any inter-node communication. Following the \textit{Inter} phase, the updates from workers on other nodes
can be quickly propagated
among local workers in this phase.

The two phases can be realized easily with GD 
operations. 
Specifically, two groups are inserted
to the GB of each worker.
Each group is generated by a GD, one
is mainly among Head Workers in different nodes (the \textit{Inter} phase), the other
is purely among local workers in the same node (the \textit{Intra} phase). 
An example can be seen in Figure \ref{fig:inter_intra}.

It is worth noting that the proposed 
{\em Inter-Intra Synchronization} is not the same
as hierarchical All-Reduce\cite{choblueconnectHierarchy},
which is mathematically equivalent to All-Reduce among all workers in one step with acceleration brought by the hierarchical architecture. 
After an All-Reduce,
all workers end up with the same weight. 
Differently, {\em Inter-Intra} synchronization strategy spreads multiple partial updates through P-Reduce in 
an architecture-aware and 
controlled manner. 
Thus, workers end up with {\em different} weights after the synchronization.


\subsection{Tolerating Slowdown}

The mechanisms proposed so far are mainly 
effective in homogeneous execution environment but 
do not help with slowdown situations. Slow workers involved in groups can block 
the current and other groups
as mentioned earlier.

\begin{figure}
    \centering
    \includegraphics[width=0.96\linewidth]{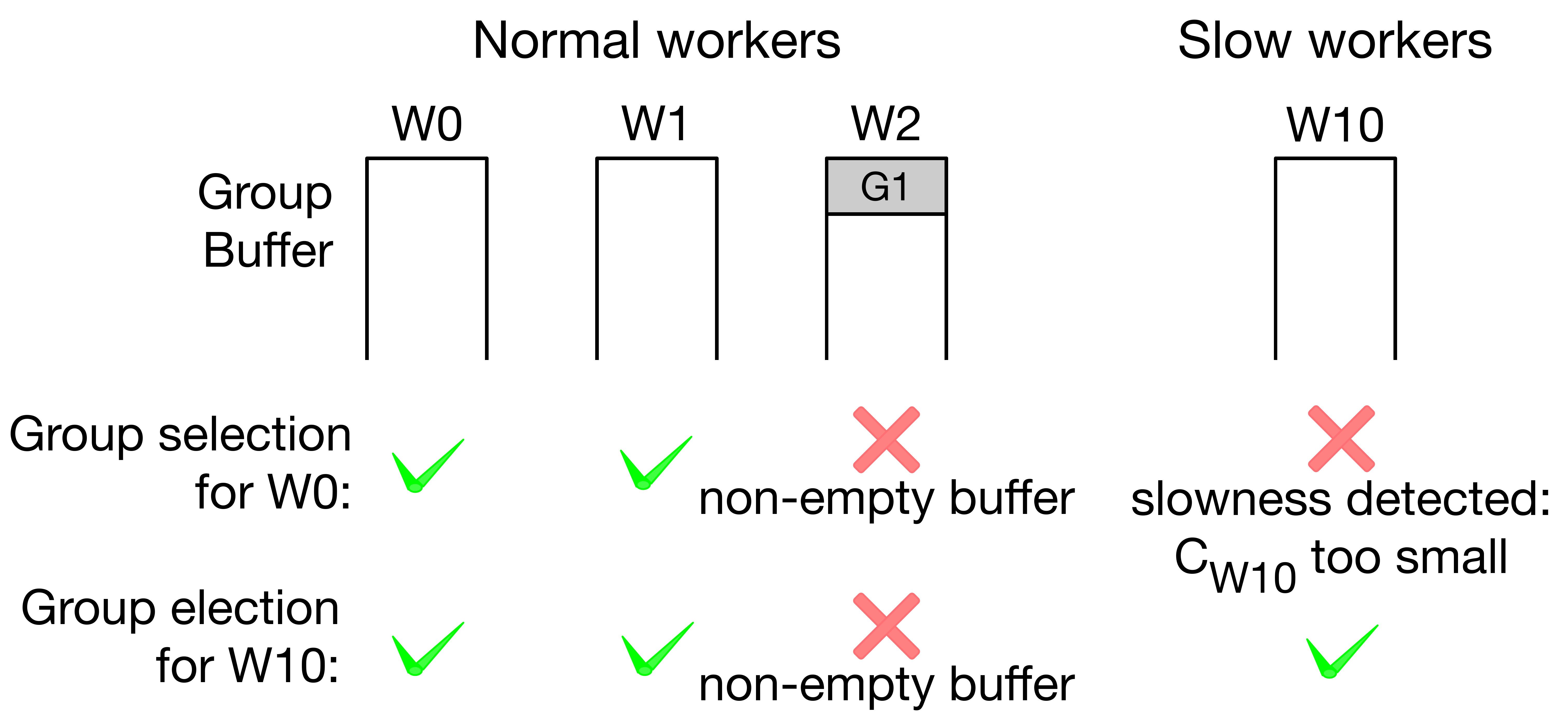} 
     \vspace{-1mm}
    \caption{Tolerating slow workers}
    \vspace{-1mm}
    \label{fig:slowdown_tolerance}
\end{figure}

We propose a simple solution by 
keeping track of execution information in GG.
Specifically, an
additional counter for each worker is placed in GG, 
which records how many times the worker requires a group. When a worker is significantly slower than other workers, the value of its counter should be also much smaller than the average. 
As a {\em GD} starts when a worker with an empty GB requests a group,
an additional rule is added to filter the workers 
who can get a group in the division:
the worker's counter, $c_w$, should be not significantly smaller than the initiator's counter, $c_i$, i.e., $c_i - c_w < C_{thres}$,
where $C_{thres}$ is a constant that can be adjusted.


This filter works as follows. When a fast worker initiates a GD, only fast workers are assigned to groups, avoiding the problem of being blocked by slow workers. When a slow worker initiates a division, some faster workers may be involved to synchronize with it. But the selected workers have empty buffers 
as defined in GD operation.
So, neither the fast workers or the slow worker needs to wait for a long time for synchronization. By the filter rule, the effect of slow workers is minimized.
\section{Implementation}
\label{sec:impl}
We implement the proposed algorithms 
and protocols using
TensorFlow and its extensions. 
Specifically, 
\projectname is implemented as customized operators of TensorFlow.

\subsection{Partial All-Reduce}

Partial All-Reduce is implemented as a GPU TensorFlow Operator. It takes the variables and the group as input tensor, and outputs a new tensor representing the result of synchronization. NCCL~\cite{jeaugey2017nccl} is used to execute All-Reduce, and MPI is used to help create NCCL communicator. 
We use a simple but effective strategy to
concatenate all weights into one tensor. 
Specifically, 
all weights are flattened and concatenated 
into one tensor for faster P-Reduce, 
and are separated and reshaped after the P-Reduce operation.

In NCCL, the upper bound of existing communicators is $64$. But it is inefficient to destroy all the communicators after use. To save the time of creating communicators, a distributed cache for communicators is used, which provides consistent presence of communicators. It does not remove cached items, but simply stops caching when its size exceeds a threshold.  

\subsection{Group Generator}
Group Generator is a centralized controller among
all workers. It requires low latency remote function call. RPC is used in this scenario. 
The server is a light-weight Python program implemented by gRPC Python package. C++ is used in the core of the algorithms. It can be started and killed easily.

The client is wrapped up as another TensorFlow Python Operator. One function as static scheduler is implemented according to the scheduling rules. Another function as dynamic group generator using the centralized GG also uses gRPC. We can easily switch between the methods of group generation using executing flags.

\section{Evaluation}
\label{sec:evaluation}
\subsection{Evaluation Setup}
\subsubsection{Hardware Environment}
We conduct our experiment on \textit{Maverick2} cluster of \textit{TACC} Super Computer. Maverick2 is a cluster managed by SLURM. In the GTX partition, a node is configured as shown in the table in Figure \ref{figure:gtx}.

\begin{figure}[!h]
\centering
\begin{tabular}{|l|c|}
\hline
 Model & Super Micro X10DRG-Q Motherboard \tabularnewline\hline
 Processor & 2 x Intel(R) Xeon(R) CPU E5-2620 v4 \tabularnewline\hline
 GPUs &  4 x NVidia 1080-TI GPUs \tabularnewline\hline
 Network & \makecell{Mellanox FDR Infiniband MT27500 Family\\  ConnectX-3 Adapter} \tabularnewline\hline
\end{tabular}
\caption{Configuration of a node in GTX partition, Maverick2 Cluster, TACC Super Computer\cite{maverick}}
\vspace{-2mm}
\label{figure:gtx}
\end{figure}

\subsubsection{Dataset and Model}
To test the performance of \projectname and 
compare it with other works, 
we train models on both medium and large data sets. 
First, we train VGG-16 model\cite{VGG11} on CIFAR-10\cite{Cifar10} image classification dataset. The model contains $9.23MB$ of trainable 32-bit floating-point weights. A typical training setup is selected. The learning rate of SGD optimizer is set to $0.1$, and the batch size per worker is $128$.

Additionally, ResNet50 model \cite{he2016identityResNet} is trained over ImageNet dataset \cite{ILSVRC15imagenet}, which contains $1,281,167$ images to be classified into $1,000$ classes. We aim to verify that \projectname is a valid algorithm that well converges in different tasks. The model contains $196MB$ of weights. Momentum optimizer is used with $momentum=0.9$ and $weight\_decay=10^{-4}$. The initial learning rate is $0.128$, and decays to its $0.1\times$ on epochs $30, 60, 80, 90$.
The training models are implemented using TensorFlow\cite{tensorflow2015-whitepaper}.

\subsubsection{Baseline Setup}
Parameter Server is already integrated in TensorFlow.
We implement AD-PSGD using remote variable access 
supported by the TensorFlow distributed module.
Horovod\cite{sergeev2018horovod} is adopted to set up a high-performance state-of-the-art baseline, which significantly outperforms many other implementations of All-Reduce. It is configured with  NCCL2\cite{jeaugey2017nccl} in order to achieve the best All-Reduce speed. We also tune the size of fuse buffer for better utilization of the Inifiniband network.
In all test runs, each worker occupies a whole GPU. For better affinity, we bind the process of each worker to the CPU socket it is directly attached to.
In random GG, the group size is 3. 

\subsubsection{Methodology}
We use the time it takes for the model (randomly initialized using a fixed random seed across different experiments) to achieve $loss=0.32$ as the metric of performance on VGG-16. We also inspect the loss w.r.t iteration
curve and the average duration of an iteration to analyze the effect of our optimizations.

\subsection{Interactions between Computation, Communication and Convergence}

\begin{figure}[!h]
    \centering
    \vspace{-4mm}
    \includegraphics[width=0.48\textwidth]{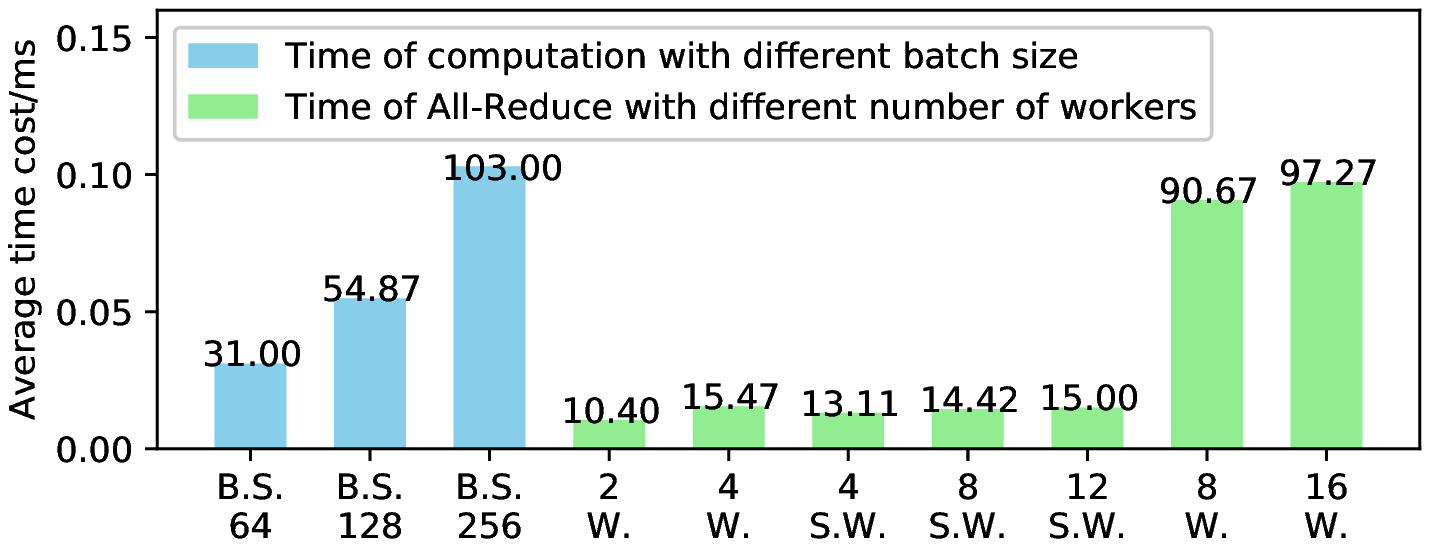}
    \footnotesize
    \emph{Notes:} B.S. means the batch size is $64, 128, 256$. W. means running $2, 4, 8, 16$ workers densely placed on $1, 1, 2, 4$ nodes. S.W. means running $4, 8, 12$ workers, one on a node, using $4, 8, 12$ nodes.
    \caption{A micro-benchmark showing the cost of different operations in computation and synchronization.}
    \vspace{-3mm}
\label{fig:components}
\end{figure}

In order to better understand how much time communication takes in deep learning training compared to computation time, we first measured the time of computation with different batch sizes and time of communication with different settings\footnote{Size of weight to be synchronized is independent of batch size}. Figure \ref{fig:components} shows the time comparisons. Because of better utilization of SIMD devices, the computation is slightly more efficient when the batch size is larger. 
Interestingly, All-Reduce among workers within a single node or workers separately placed across 
different nodes are significantly faster than 
having multiple nodes with each running multiple workers.


\begin{figure}[!h]
    \centering
    \includegraphics[width=0.48\textwidth]{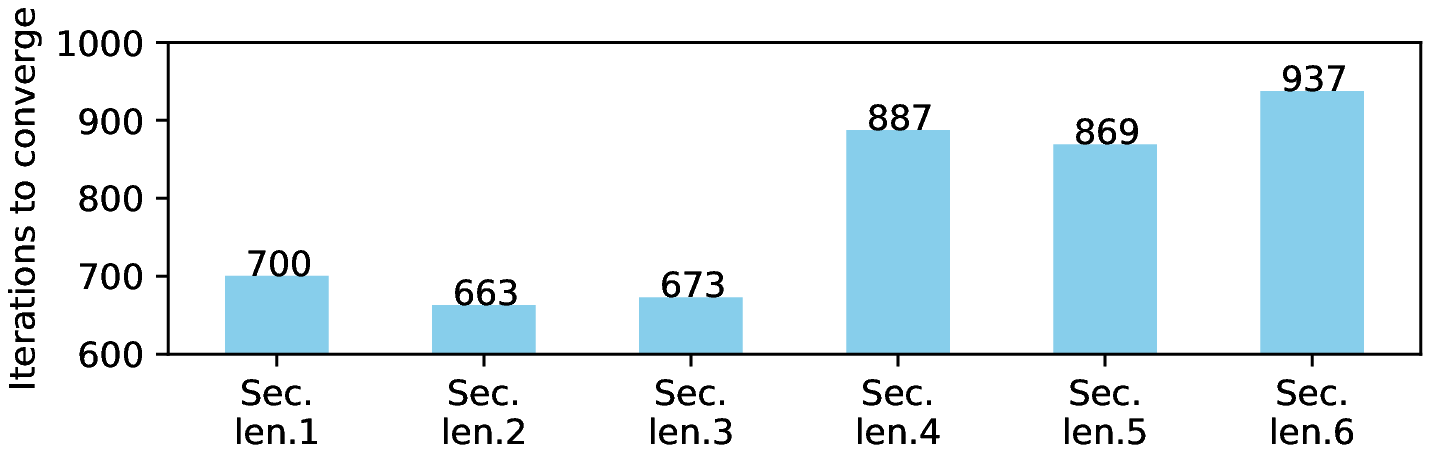}
    \footnotesize
    \emph{Notes:} The frequency of communication is controlled by a hyper-parameter {\em Section Length}, -- \# of iterations between two synchronizations. 
    \caption{Effects of reducing synchronization}
    \vspace{-2mm}
    \label{fig:seclen}
\end{figure}

Although reducing communication by lowering 
synchronization frequency 
can increase the throughput of training, 
it becomes harder to converge. Figure \ref{fig:seclen} presents a simple experiment to show that the number of iterations needed to converge increases as communication frequency gets lower. To get the best performance of convergence time, 
setting a proper level of synchronization intensity is necessary.
This result shows that we cannot simply
improve AD-PSGD by enlarging the amount of computation 
between synchronizations. 

\subsection{Speedup in Homogeneous Environment}
\label{sec:homospeedup}
\begin{figure}[!h]
    \centering
    \vspace{-3mm}
    \includegraphics[width=0.48\textwidth]{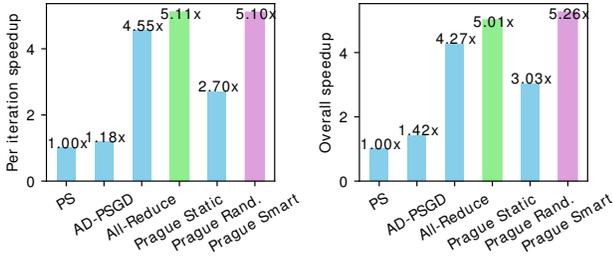}
    \vspace{-6mm}
    \caption{Per-iteration speedup and overall speedup}
    \vspace{-2mm}
    \label{fig:homospeedup}
\end{figure}

In a homogeneous environment with $16$ workers on $4$ nodes, VGG-16 trained over CIFAR-10 is used to compare \projectname with {\em different ways of group generation} against Parameter Server, All-Reduce and AD-PSGD.
The per-iteration speedup and convergence time speedup is shown in Figure \ref{fig:homospeedup}. \projectname is much faster than Parameter Server and the original AD-PSGD. All-Reduce is also much faster than these two baselines, due to the high throughput provided by Horovod. However, \projectname with both static scheduler and smart GG even outperform All-Reduce thanks to its smaller synchronization groups and architecture-aware
scheduling. 

\begin{figure}[!h]
    \centering
    \vspace{-3mm}
    \includegraphics[width=0.48\textwidth]{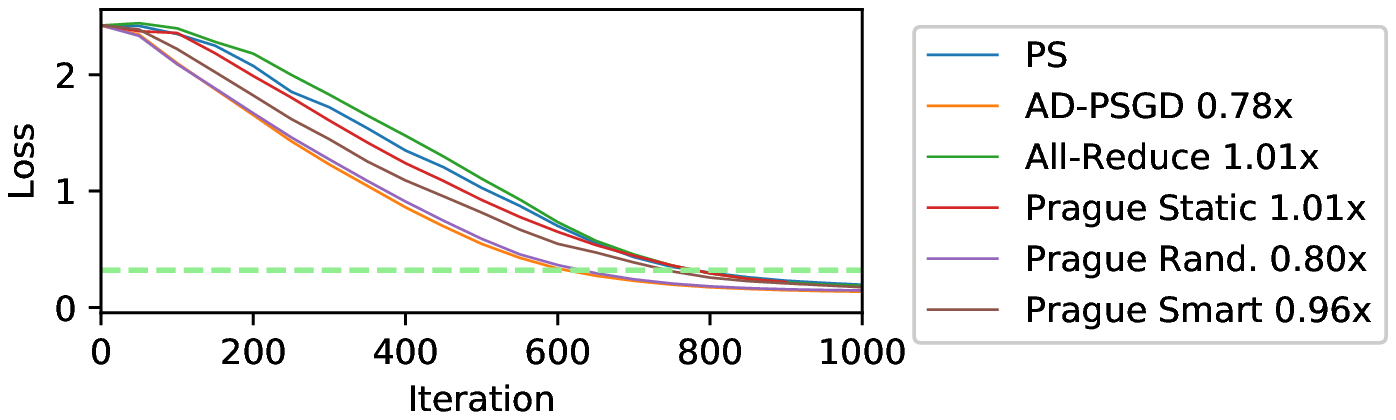}
    \footnotesize
    {\emph Notes:} The speedup in the figure means the number of iterations to converge compared to Parameter Server.
    \caption{Convergence curve in terms of number of iterations for corresponding algorithms in Figure \ref{fig:homospeedup}}
    \vspace{-2mm}
    \label{fig:homolossstep}
\end{figure}

Shown in Figure~\ref{fig:homolossstep}, AD-PSGD has better convergence speed in terms of number of iterations. All-Reduce is mathematically equivalent to Parameter Server. They are slightly different due to random sampling and competition in synchronization.
\projectname with static scheduler has similar convergence speed as Parameter Server, but it gains speedup from its higher throughput. 
We see that the number of iterations in random GG
is less than smart GG, which is 
smaller than static scheduling. 
This is due to the decreasing amount
of randomness from random GG to smart GG and to
static scheduling. 


These results further demonstrate the 
trade-offs between execution efficiency and 
statistical efficiency~\cite{stanford_chris_re}.
Although AD-PSGD needs fewer iterations to 
converge to the same error, the execution time
of each iteration is seriously affected by
the synchronization overhead, shown in 
Figure~\ref{fig:ratio} (b). 
\projectname successfully explores this 
trade-off by slightly sacrificing 
statistical efficiency, i.e.,
running more iterations (0.96x vs. 0.78x), --- 
mainly caused by 
the reduced randomness, to gain significant 
speedup in per iteration execution time (5.10x vs. 1.18x) 
and eventually lead to overall execution time speedup
(5.26x vs. 1.42x). 

\subsection{Heterogeneity Tolerance}

\begin{figure}[!h]
    \centering
    \vspace{-4mm}
    \includegraphics[width=0.48\textwidth]{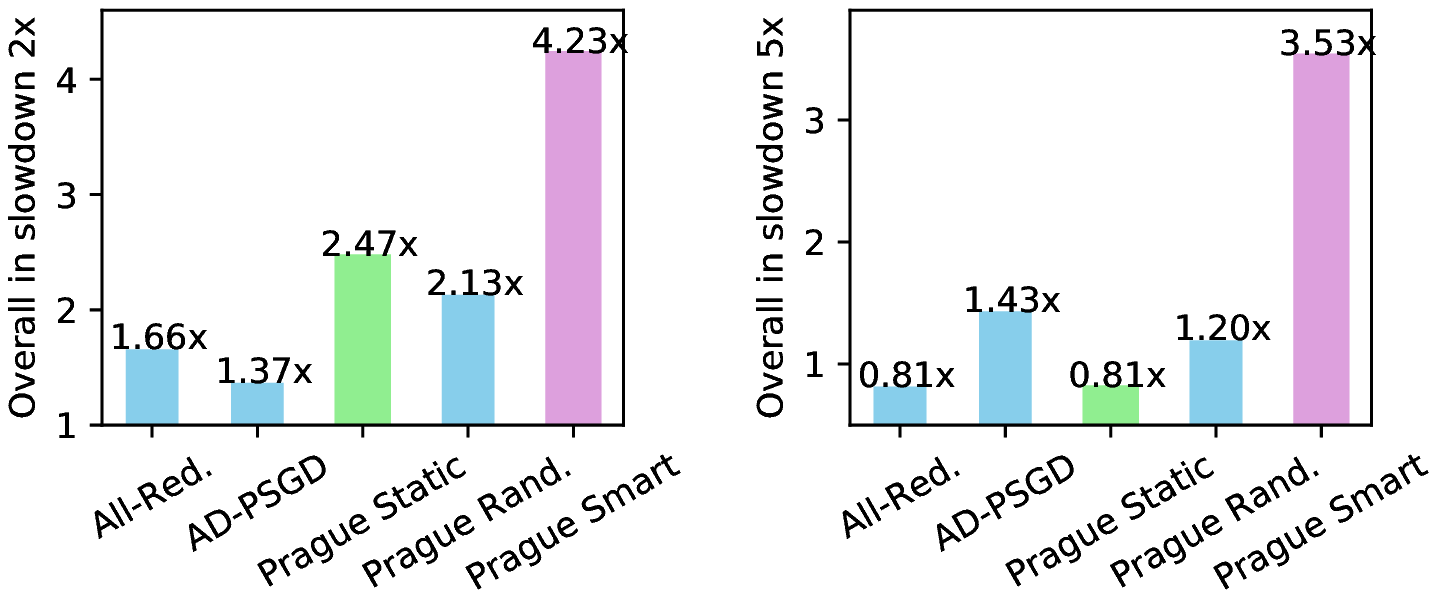}
    \footnotesize
    \emph{Notes:} The baseline is still Parameter Server without slowdown in Figure \ref{fig:homospeedup} in convenience for comparing.
    \caption{Overall speedup of All-Reduce, \projectname with static scheduler and \projectname with random and smart GG in heterogeneous environment (2x or 5x slowdown on one worker).}
    \vspace{-3mm}
    \label{fig:slowdown}
\end{figure}

One of the key advantages of \projectname is better tolerance of heterogeneity. Based on the same setup in section \ref{sec:homospeedup}, heterogeneity is simulated by adding $2$ or $5$ times the normal iteration time of sleep every iteration on one specific worker, the slow worker.
The result is shown in Figure \ref{fig:slowdown}. 
In terms of the capability to tolerate slowdown, 
experiment results of 2x slowdown show that:
(1) random GG (3.03x vs. 2.13x) is slightly
worse than AD-PSGD 
(1.42x vs. 1.37x), but it is much faster due to 
more efficient P-Reduce as the synchronization primitive; 
(2) smart GG (5.26x vs. 4.23x) is better than random GG
(3.03x vs. 2.13x); and 
(3) while both suffer from more slowdown, 
\projectname static (5.01x vs. 2.47x) is still 
considerably better than All-Reduce (4.27x vs. 1.66x).
We also see that with 2x slowdown, 
All-Reduce is still faster than AD-PSGD although 
much slower than itself in homogeneous setting.
With 5x slowdown, All-Reduce can only achieve 
a little more than half of the performance 
in AD-PSGD. We see that random GG is slightly 
slower than AD-PSGD, this is because the larger
group size (3) in \projectname can increase the 
chance of conflicts. Nevertheless, 
smart GG outperforms AD-PSGD with a large margin.



\subsection{Validation on Large Model and Dataset}
This section shows the training performance of 
ResNet-50 on ImageNet by 
running only $10$ hours of training on $8$ nodes with $32$ workers for each algorithm.
We conduct experiment in this manner to avoid affecting other experiments on the cluster, as TACC Super Computer is shared by thousands of researchers. 

\begin{figure}[!h]
    \centering
    \begin{tabular}{|l|c|c|c|}
    \hline
    Algorithm & Total iterations & \makecell{Top 1\\Accuracy} & \makecell{Top 5\\Accuracy} \tabularnewline\hline
    All-Reduce & 55800 & 66.83\% & 84.81\% \tabularnewline
    AD-PSGD & 32100 & 58.28\% & 78.00\% \tabularnewline
    Prague Static & 58200 & 63.79\% & 82.38\% \tabularnewline
    Prague Smart & 56800 & 64.21\% & 82.78\% \tabularnewline
    \hline
    \end{tabular}

    \includegraphics[width=0.48\textwidth]{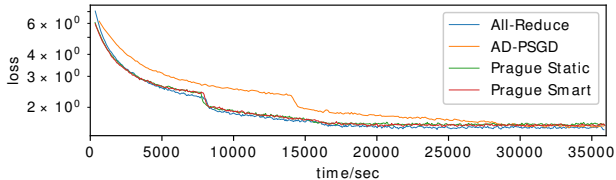}
    \vspace{-6mm}
    \caption{Iterations trained, final training accuracy of different algorithms after training for $10$ hours, and loss curve during the $10$ hours.}
    \vspace{-1mm}
    \label{fig:imagenetcurve}
\end{figure}

The training accuracy and the loss curves for the
10-hour executions are shown in Figure \ref{fig:imagenetcurve}. Please note the execution
environment is homogeneous without slower workers. 
We see that 
All-Reduce performs the best in this case, followed by \projectname with smart GG.
AD-PSGD suffers from throughput issue. In ResNet-50 over ImageNet, the upper bound of effective batch size is very large. Therefore, although we make our best effort to enlarge the batch size, All-Reduce obtains much bigger convergence advantage numerically, while \projectname can train more iterations using the same time. The smart GG performs better than static scheduler because it has more randomness in synchronization. Observing from the loss curve, \projectname still has competitive convergence speed compared with the state-of-the-art approach, All-Reduce, on large data sets. 

\section{Conclusion}

In this paper, we propose {\em \projectname},
a high-performance heterogeneity-aware 
asynchronous decentralized training approach.
To reduce 
synchronization cost, we propose a novel communication primitive,
Partial All-Reduce, that allows 
a large group of workers to synchronize quickly.
To reduce
synchronization conflict, we propose static 
group scheduling in homogeneous environment
and simple techniques (Group Buffer and Group 
Division) to avoid conflicts with slightly 
reduced randomness. 
Our experiments show that in homogeneous environment, \projectname is $1.1\times$ faster than the state-of-the-art implementation of All-Reduce, and is $5.1\times$ faster than Parameter Server and $4.3\times$ faster than AD-PSGD. In a heterogeneous setting,
\projectname shows $2\times$ speedup over All-Reduce, and still obtains $3\times$ speedup over AD-PSGD.


\bibliographystyle{plain}
\bibliography{references}

\end{document}